\documentclass{emulateapj}
\usepackage{epsfig,natbib}
\usepackage{graphicx}
\newcommand{\ms}{\mbox{m\,s$^{-1}~$}}

\newcommand{\mse}{\mbox{m\,s$^{-1}$}}
\newcommand{\msy}{\mbox{m\,s$^{-1}$\,yr$^{-1}~$}}

\newcommand{\mjup}{M$_{\rm JUP}~$}

\newcommand{\mearth}{$M_\earth$~}
\newcommand{\mearthe}{$M_\earth$}
\newcommand{\rearth}{$R_\earth$~}
\newcommand{\rearthe}{$R_\earth$}

\newcommand{\msini}{$M \sin i~$}
\newcommand{\msinie}{$M \sin i$}

\newcommand{\feh}{\ensuremath{[\mbox{Fe}/\mbox{H}]}}
\newcommand{\rphk}{\ensuremath{R'_{\mbox{\scriptsize HK}}}}
\newcommand{\lrphk}{\ensuremath{\log{\rphk}}}
\newcommand{\caii}{\ion{Ca}{2} H \& K}
\newcommand{\caiih}{\ion{Ca}{2} H}

\shortauthors{Howard {et~al.}}
\shorttitle{A Super-Earth Orbiting HD\,156668}
\submitted{Submitted to ApJ}
\begin{document}
\pagenumbering{arabic}


\title{The NASA-UC Eta-Earth Program: \\
         II. A Planet Orbiting HD\,156668 with a Minimum Mass of Four Earth Masses\altaffilmark{1}}
\author{
Andrew W.\ Howard\altaffilmark{2,3}, 
John Asher Johnson\altaffilmark{4}, 
Geoffrey W.\ Marcy\altaffilmark{2}, 
Debra A.\ Fischer\altaffilmark{5}, \\
Jason T.\ Wright\altaffilmark{6}, 
Gregory W.\ Henry\altaffilmark{7},
Howard Isaacson\altaffilmark{2}, \\
Jeff A.\ Valenti\altaffilmark{8},
Jay Anderson\altaffilmark{8}, and
Nikolai E.\ Piskunov\altaffilmark{9} 
}
\altaffiltext{1}{Based on observations obtained at the W.\,M.\,Keck Observatory, 
                      which is operated jointly by the University of California and the 
                      California Institute of Technology.  Keck time has been granted by both 
                      NASA and the University of California.} 
\altaffiltext{2}{Department of Astronomy, University of California, Berkeley, CA 94720-3411, USA} 
\altaffiltext{3}{Townes Fellow, Space Sciences Laboratory, University of California, 
                        Berkeley, CA 94720-7450 USA; howard@astro.berkeley.edu}
\altaffiltext{4}{Department of Astrophysics, California Institute of Technology, MC 249-17, Pasadena, CA 91125, USA}
\altaffiltext{5}{Department of Astronomy, Yale University, New Haven, CT 06511, USA}
\altaffiltext{6}{The Pennsylvania State University, University Park, PA 16802}
\altaffiltext{7}{Center of Excellence in Information Systems, Tennessee State University, 
                        3500 John A.\ Merritt Blvd., Box 9501, Nashville, TN 37209, USA}
\altaffiltext{8}{Space Telescope Science Institute, 3700 San Martin Dr., Baltimore, MD 21218, USA}
\altaffiltext{9}{Department of Astronomy and Space Physics, Uppsala University, 
                        Box 515, 751 20 Uppsala, Sweden}

\begin{abstract}
We report the discovery of HD\,156668\,b, an extrasolar planet with a minimum mass of $M_P\sin i$\,=\,4.15\,\mearthe.
This planet was discovered through Keplerian modeling of precise radial velocities from Keck-HIRES 
and is the second super-Earth to emerge from the NASA-UC Eta-Earth Survey.
The best-fit orbit is consistent with circular and has a period of $P$\,=\,4.6455\,d.
The Doppler semi-amplitude of this planet, $K$\,=\,1.89\,\mse, is among the lowest ever detected, 
on par with the detection of GJ\,581\,e using HARPS. 
A longer period ($P$\,$\approx$\,2.3\,yr), low-amplitude signal of unknown origin was also detected 
in the radial velocities and was filtered out of the data while fitting the short-period planet.  
Additional data are required to determine if the long-period signal is due to a second planet, stellar activity, or another source.
Photometric observations using the Automated Photometric Telescopes at Fairborn Observatory show that 
HD\,156668 (an old, quiet K3 dwarf) is photometrically constant over the radial velocity period to 
0.1\,mmag, supporting the existence of the planet.
No transits were detected down to a photometric limit of $\sim$3\,mmag, 
ruling out transiting planets dominated by extremely bloated atmospheres, 
but not precluding a transiting solid/liquid planet with a modest atmosphere.
\end{abstract}


\section{Introduction}
\label{sec:intro}

The search for low-mass planets is driven by a desire to 
observationally study the full range of planetary systems in order 
to better understand their formation, dynamics, composition, and diversity.  
We also seek Earth-like worlds of terrestrial composition as a goal in itself 
and as targets for future transit and imaging observations.  
This search has taken several leaps forward recently because of instrumental improvements.
The precision of radial velocity (RV) measurements with 
Keck-HIRES by the California Planet Survey (CPS) group \citep{Howard09a}, 
HARPS \citep{Lovis06}, and the AAT \citep{OToole09} has now reached 1\,\ms or better and has led 
to the discovery of several super-Earths around nearby, bright stars.  
Ground-based transit surveys such as MEarth \citep{Charbonneau09}  and HATNet \citep{Bakos09} 
have made important discoveries of transiting low-mass planets.
Microlensing searches have detected two super-Earths orbiting distant stars \citep{Beaulieu06,Bennett08} 
and the statistics of microlensing detections suggest than cold Neptunes are a factor of three 
more common that cold Jupiters \citep{Sumi09}.
From space, \textit{CoRoT} has found a system with two super-Earths (one of which transits; \citealt{Leger09}) 
and \textit{Kepler} is poised to detect true Earth analogues in 1\,AU orbits using transit photometry 
with a precision of 20\,ppm in 6.5\,hr \citep{Borucki09}.
In the next decade, \textit{SIM Lite} \citep{Unwin08} will astrometrically characterize essentially all planets 
down to Earth mass orbiting $\sim$100 nearby stars, 
as well as the more massive planets orbiting $\sim$1000 stars.  

The Eta-Earth Survey plays a unique role in the study of low-mass exoplanets.
The population of 230 GKM stars in the survey is nearly free of statistical bias since 
the stars were not chosen based on their likelihood of harboring a planet, but rather on proximity, brightness, and chromospheric activity.  
Each star is observed at least 20 times, insuring a minimum detectability threshold.
Thus, the distributions of planet detections and non-detections from the Eta-Earth Survey  
will yield a wealth of information about the efficiency and mechanisms of planet formation 
as well as the range of subsequent dynamical histories. 
The 20 survey observations per Eta-Earth Survey target are nearly complete 
and we are aggressively re-observing several promising candidate low-mass planets.

Current theories of planet formation \citep{Ida04a,Mordasini09} 
are consistent with the measured distributions of massive planets (Saturn mass and above), 
but their predictions for the abundance and properties of low-mass planets are only now being observationally tested.
In particular, they predict a dearth of planets below roughly Saturn mass in orbits inside the ice line.  
Such planets are predicted to be rare because once a planet core grows by planetesimal accretion
to a of threshold mass of $\sim$10\,\mearthe, it undergoes runaway gas accretion and becomes a gas giant.
While these theories consistently predict such a ``planet desert'', 
they differ in the contours of the desert in $M$ and $a$ as well as 
whether the planets below the critical core mass survive Type I inward migration in significant numbers.

In this context we announce the discovery of HD\,156668\,b, a super-Earth planet with minimum mass   
$M_P\sin i$\,=\,4.15\,\mearth and an orbital period of $P$\,=\,4.6455\,d.
This is the second super-Earth (\msinie\,$<$\,10\,\mearthe) to emerge from Keck observations 
explicitly for the NASA-UC Eta-Earth Survey, the first being HD\,7924\,b \citep{Howard09a}.  
The remainder of this paper is structured as follows.  
We describe the host star properties in \S\,\ref{sec:props}.  
The spectroscopic observations and their Doppler reduction are described in \S\,\ref{sec:obs}.
In \S\,\ref{sec:orbital}, we describe the detection of the $P$\,=\,4.6455\,d orbit of HD\,156668\,b, 
and the high-pass filtering of the RV data that was necessary to obtain good estimates of the 
orbital parameters.  
In \S\,\ref{sec:null} we carefully consider the null hypothesis---the non-existence of HD\,156668\,b---using 
$S_\mathrm{HK}$ measurements, photometric observations, and FAP analyses.
We summarize and discuss the implications of this discovery in \S\,\ref{sec:discussion}.

\section{Stellar Properties}
\label{sec:props}

HD\,156668 (HIP\,84607) is a K3 dwarf \citep{Gray03} whose properties 
are summarized in Table \ref{tab:stellar_params}.
It is nearby ($d$\,=\,24.5\,pc; \citealt{vanLeeuwen07}) and relatively bright ($V$\,=\,8.424; \citealt{tycho_cat00}).
With $M_V$\,=\,6.480 and $B-V$\,=\,1.015, the star lies near
the $Hipparcos$ average main sequence as defined by \citet{Wright05}.

\begin{deluxetable}{lc}
\tabletypesize{\footnotesize}
\tablecaption{Stellar Properties of HD\,156668
\label{tab:stellar_params}}
\tablewidth{0pt}
\tablehead{
  \colhead{Parameter}   & 
  \colhead{Value} 
}
\startdata
Spectral type ~~~~~~~~~~~~~~~~& K3\,V\\
$M_V$ & 6.480\\
$B-V$ & 1.015\\
$V$   & 8.424\\
$J$   & 6.593\\
$H$   & 6.117\\
$K$   & 6.004\\
Distance (pc) & 24.5\,$\pm$\,0.5\\
\feh& $+0.05$\,$\pm$\,0.06\\
$T_\mathrm{eff}$ (K) &  4850\,$\pm$\,88\\
$v$\,sin\,$i$ (km\,s$^{-1}$) & 0.50\,$\pm$\,1.0 \\
log\,$g$ & 4.598\,$\pm$\,0.12\\
$L_{\star}$ ($L_{\sun}$) & 0.230\,$\pm$\,0.018\\
$M_{\star}$ ($M_{\sun}$) & 0.772\,$\pm$\,0.020\\
$R_{\star}$ ($R_{\sun}$) & 0.720\,$\pm$\,0.013\\
Age (Gyr) & 8.6\,$\pm$\,4.8\\
\lrphk & $-4.98$\\
$S_\mathrm{HK}$ & 0.23\\
$P_\mathrm{rot}$ (days) & 51.5\\
$\sigma_\mathrm{phot}$ (mag) & $\lesssim$\,0.002\\
\enddata
\end{deluxetable}

Using the SME (Spectroscopy Made Easy) LTE spectral synthesis code \citep{Valenti05}, 
we analyzed two high-resolution, iodine-free Keck-HIRES spectra of HD\,156668 and found 
the effective temperature, surface gravity, projected rotational velocity, 
and iron abundance ratio listed in Table \ref{tab:stellar_params}.  
The errors on these four quantities have been conservatively doubled 
from the formal SME parameter uncertainties 
because the stellar atmosphere models are less certain at low $T_\mathrm{eff}$ 
(the SME catalog \citep{Valenti05} models stars down to $T_\mathrm{eff}$ = 4800\,K).
We refined the above parameters and derived stellar mass, radius, and luminosity 
from SME and interpolated Yonsei-Yale (Y$^2$) isochrones \citep{Demarque04} 
using an iterative process that self-consistently ties together 
the SME and Y$^2$ values of log\,$g$ \citep{Valenti09}.
HD\,156668 appears to be a typical middle-aged K dwarf.  
Its slightly super-solar iron abundance of \feh\,=\,$+0.05$\,$\pm$\,0.06 
is consistent with its location near the average main sequence.

\begin{figure}
\epsscale{1.15}
\plotone{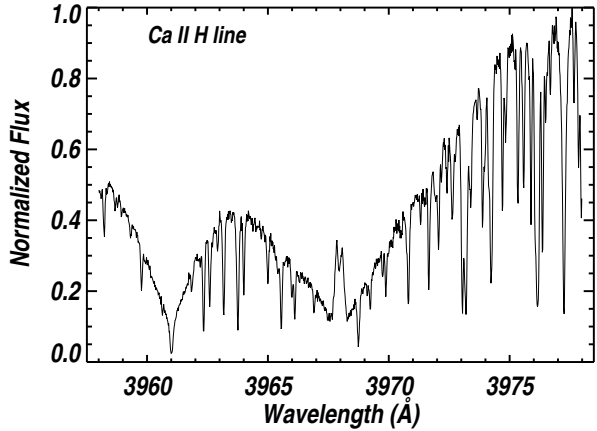}
\caption{\caiih\ line for HD\,156668 from a Keck/HIRES spectrum.  
The line core emission near 3968\,\AA\ indicates modest chromospheric activity.}
\label{fig:caii}
\end{figure}

Measurements of the cores of the \caii\ lines show that 
HD\,156668 has modest chromospheric activity (Fig.\ \ref{fig:caii}).
We measured the chromospheric activity indices 
$S_{\mathrm{HK}}$\,=\,0.23 and \lrphk\,=\,$-$4.98 using the method described \cite{Wright04} and \cite{Isaacson09}.
The full set of $S_{\mathrm{HK}}$ measurements for all observations of HD\,156668 
does not show a periodicity near the planet's orbital of 4.6455\,d 
(see \S\,\ref{sec:chromospheric} for additional discussion).

We estimate a rotation period $P_{\mathrm{rot}}$\,$\sim$\,48\,d using \rphk\ and $B-V$ calibration \citep{Noyes84}, 
which is consistent with the 51.5\,d period measured by APT photometry (see \S\,\ref{sec:photometry}).
Following \citet{Wright05}, and based on the values of $S_{\mathrm{HK}}$, $M_V$, and $B-V$, 
we estimate an RV jitter of 1.5\,\mse.  
This empirical estimate is based on an ensemble of stars with similar characteristics 
and accounts for RV variability due to 
rotational modulation of stellar surface features, stellar pulsation, undetected planets, 
and uncorrected systematic errors in the velocity reduction \citep{Saar98,Wright05}.  
As explained in \S\,\ref{sec:orbital}, jitter is added in quadrature to the 
RV measurement uncertainties for Keplerian fitting.

HD\,156668 has several important characteristics that make it a nearly ideal RV target star.  
Like other old, chromospherically quiet stars with spectral types from late G to early K, 
HD\,156668 appears to be near the 
minimum of astrophysical jitter arising from acoustic oscillations, granulation, and photospheric activity
(see \S\,\ref{sec:chromospheric} and Figure~\ref{fig:rphk}).  
The star is relatively bright, yielding high signal-to-noise spectra in $\sim$4--5 minutes per observation.

\section{HIRES Observations and Doppler Reduction}
\label{sec:obs}

We observed HD\,156668 with the HIRES echelle spectrometer \citep{Vogt94} 
on the 10-m Keck I telescope.  The 107 observations span five years (2005--2009) 
with high-cadence observations---clusters of observations on 6--12 
consecutive nights---beginning in 2007.  
All observations were made with an iodine cell mounted directly in front of the 
spectrometer entrance slit.  The dense set of molecular absorption lines imprinted 
on the stellar spectra provide a robust wavelength fiducial 
against which Doppler shifts are measured, 
as well as strong constraints on the shape of the spectrometer instrumental profile at 
the time of each observation \citep{Marcy92,Valenti95}.

We measured the Doppler shift from each star-times-iodine spectrum using a 
modeling procedure modified from the method described by \citet{Butler96b}.  
The most significant modification is the way we model the
intrinsic stellar spectrum, which serves as a reference point for the relative Doppler
shift measurements for each observation.  Butler et al.\ use a version of the
\citet{Jansson95} deconvolution algorithm to remove the spectrometer's instrumental profile
from an iodine-free template spectrum.  We instead use a new deconvolution
algorithm developed by one of us (J.\,A.\,J.) that employs a more effective regularization
scheme, which results in significantly less noise amplification and improved
Doppler precision. 

\begin{figure}
\epsscale{1.15}
\plotone{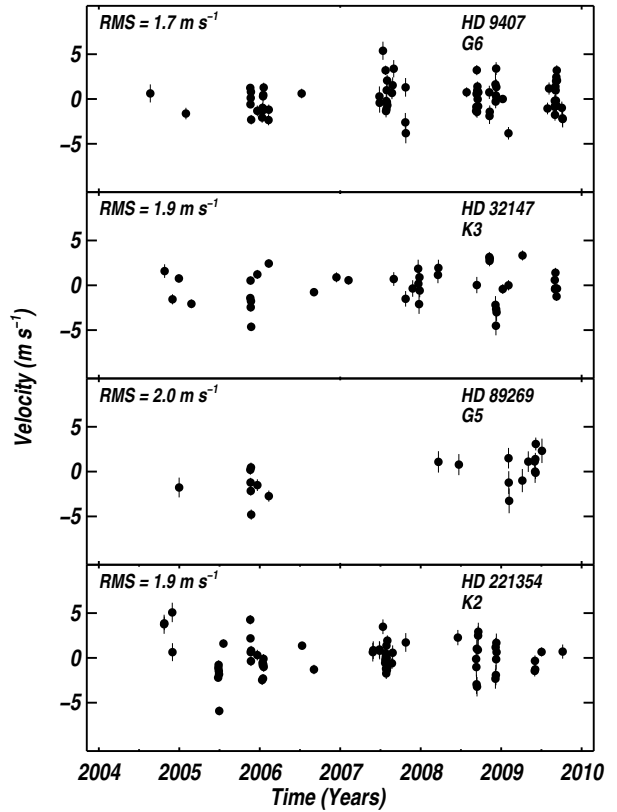}
\caption{Radial velocity time series for four stable stars  
in our Keck Doppler survey that are similar to HD\,156668.
These stars demonstrate long-term velocity stability and precision of HIRES.
The binned velocities with measurement uncertainties (but not jitter) are plotted.  
Panels are labeled with star name, spectral type, and velocity rms.}
\label{fig:standard_stars}
\end{figure}

Figure \ref{fig:standard_stars} shows RV time series for four stable stars with
characteristics similar to HD\,156668, demonstrating our measurement precision of 
$\sim$1.5--2.0\,\ms (including astrophysical, instrumental/systematic, and photon-limited errors) 
for chromospherically quiet late G/early K stars over the past 5 years. 
All of the measurements reported here were made after the HIRES CCD upgrade 
in 2004 August and do not suffer from the higher noise and systematic errors that limited the 
precision of pre-upgrade measurements to $\sim$2--3\,\ms for most stars. 

The velocities derived from the 107 observations have 
a median single measurement uncertainty of 1.01\,\mse.  
This uncertainty is the weighted standard deviation of the mean of the velocity measured 
from each of the $\sim$700 2\,$\mathrm{\AA}$ chunks in each echelle spectrum \citep{Butler96b}.
In a few cases, we made consecutive observations of HD\,156668 to reduce the 
Poisson noise from photon statistics.  
For the Keplerian orbital analysis below (\S\,\ref{sec:orbital}), 
the velocities were binned in 2\,hr intervals, 
yielding 86 measurements with an rms of 2.71\,\ms about the mean and 
a median measurement uncertainty of 0.99\,\mse.  

\begin{figure*}
\epsscale{1.15}
\plotone{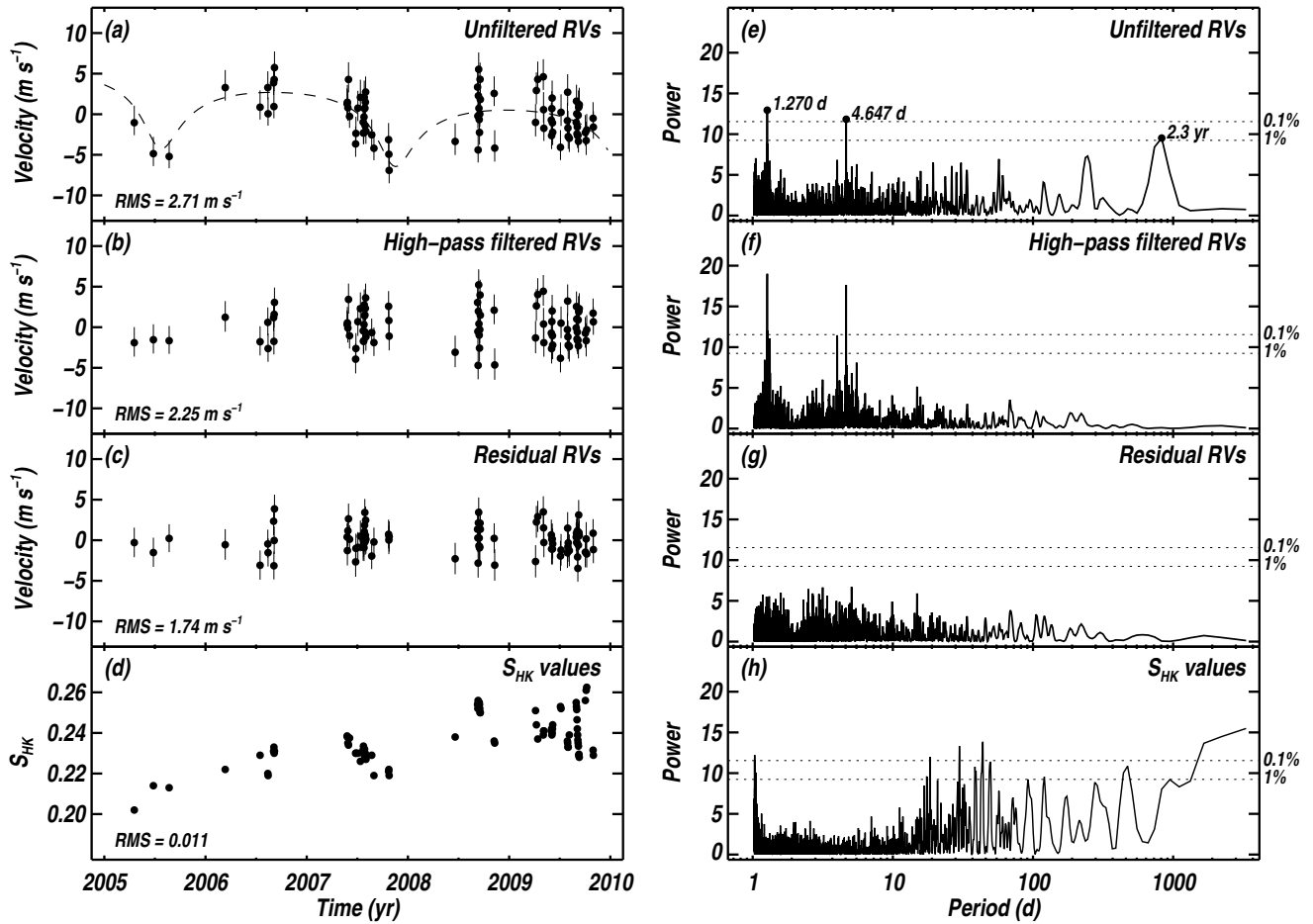}
\caption{RV measurements and $S_\mathrm{HK}$ values for HD\,156668 based on Keck-HIRES spectra.
       Each pair of panels in a given row displays the same data as a 
       time series (panels \textit{a--d}, on left) and as a periodogram (panels \textit{e--g}, on right).  
       Panels \textit{a} and \textit{e} show the unfiltered RVs from Table~\ref{tab:keck_vels} 
       with significant Fourier power at $\sim$2.3\,yr, 4.647\,d, and 1.270\,d 
       (an alias of 4.647\,d with the sidereal day length).
       Panels \textit{b} and \textit{f} show the RVs after applying a high-pass filter by subtracting 
       the dashed line model in panel \textit{a}; 
       note that the $\sim$2.3\,yr signal was almost completely removed, strengthening the 
       power at 4.647\,d and its alias.
       Panels \textit{c} and \textit{g} show the residuals of the high-pass filtered data 
       to a one-planet fit with $P$\,=\,4.6455\,d in a circular orbit.
       Panels \textit{d} and \textit{h} show values of the chromospheric index $S_\mathrm{HK}$ 
       derived from each RV measurement.  
       }
\label{fig:ts_pg}
\end{figure*}

\section{Orbital analysis}
\label{sec:orbital}

The measured radial velocities of HD\,156668 are listed in 
Table \ref{tab:keck_vels} and plotted as a time series in Figure~\ref{fig:ts_pg}a.
Before any fitting or filtering, these velocities already have a low rms, $\sigma$\,=\,2.71\,\mse.  
However, long-term trends, coherent over at least several months, are apparent by visual inspection of 
Figure~\ref{fig:ts_pg}a (see, e.g., the downward trend in 2007).  
These trends manifest themselves in a Lomb-Scargle periodogram \citep{Lomb76,Scargle82} of the data 
as long-period power concentrated near $\sim$2.3\,yr (Figure~\ref{fig:ts_pg}e).  

The RVs also show a significant periodicity near 4.647\,d with power than exceeds the 0.1\% FAP 
threshold\footnote{The false alarm probability (FAP) thresholds 
plotted as dashed horizontal lines in the periodograms in Figure~\ref{fig:ts_pg} 
refer to the probability that periodograms of random rearrangements 
of the data would exceed the specified power level.
Since periodograms only measure the power of sine wave fits to the data (i.e.\ circular orbits), 
these FAPS are less conservative than the ones described in \S\,\ref{sec:fap} that allow for 
eccentric orbits and use the $\Delta\chi_{\nu}^2$ statistic.}.
A second short-period peak is visible at 1.270\,d, but this peak is a stroboscopic alias 
of the 4.647\,d signal with the sidereal day length (see \S\,\ref{sec:alias}).
We attempted to fit the radial velocities with a single-planet Keplerian orbital solution seeded 
with $P$\,=\,4.647\,d using a partially-linearized, least-squares fitting procedure \citep{Wright09b}.
Each velocity measurement was assigned a weight 
constructed from the quadrature sum of the measurement uncertainty 
(listed in Table \ref{tab:keck_vels}) and a stellar jitter term (1.5\,\mse).  
The fitting routine converged on a robust solution with $P$\,=\,4.6455\,d, 
$e$\,=\,0.39, and $K$\,=\,2.28\,\mse.
This model gives $\sigma$\,=\,2.33\,\ms and $\chi_{\nu}$\,=\,1.70, 
a significant improvement over a linear model to the data.

We also tried fitting the long-period signal with a single-planet Keplerian model.  
We seeded the fitting routine with a range of periods and found a best-fit solution 
with $P$\,=\,791\,d (2.17\,yr) and a poorly constrained eccentricity 
when that parameter is allowed to float.  
The improvements in $\sigma$ and $\chi_{\nu}$ for this model 
are comparable to those from the $P$\,=\,4.6455\,d model described above.
However, the poorly constrained and high eccentricity of the fit and the uneven phase coverage render 
the interpretation of this signal suspect.  
We cannot be sure if it is due to a planet in Keplerian motion, 
an astrophysical signal such as chromospheric activity masquerading as a RV change of the star, 
or some other effect.

Whatever the \textit{source} of the long-period signal, its \textit{effect} is to obscure the short-period signal.  
We considered several ways to high-pass filter the RVs
to isolate the short-period signal.  
A key requirement of such a filtering process is that it leave the short-period signal untouched.  
We concluded that fitting the data to a two-planet model plus a linear trend 
would robustly separate the long and short-period signals, 
allowing for accurate parameter estimation for the short-period signal and a fair assessment of its statistical significance.  
The fitting routine converged on a robust solution with Keplerian parameters for the inner planet 
listed in Table~\ref{tab:orbital_params}.  We adopt these parameters for HD\,156668\,b.   
The ``outer planet'' in this model has $P$\,=\,2.31\,yr, $e$\,=\,0.48, $K$\,=\,4.05\,\mse, $\omega$\,=\,178\,deg, 
$T_{\mathrm{p}}$\,=\,2,454,414.22, and $dv/dt$\,=\,0.94\,\msy as shown by the dashed line in Figure~\ref{fig:ts_pg}.
We emphasize that the outer planet in this fit is only a convenient model for the data, 
effectively serving as a high-pass filter to isolate the signal of the inner planet.  
Determining whether the long-period signal represents a planet will require 
additional RV measurements and diagnostic data.

\begin{deluxetable}{lc}
\tabletypesize{\footnotesize}
\tablecaption{Orbital Solution for HD\,156668\,b
\label{tab:orbital_params}}
\tablewidth{0pt}
\tablehead{
\colhead{Parameter}   & \colhead{Value} 
}
\startdata
$P$ (days)     & 4.6455 $\pm$ 0.0011 \\
$T_c$ (JD -- 2,440,000) & 14718.57 $\pm$ 0.11 \\
$e$                     & $\equiv$0.0\\
$K$ (m\,s$^{-1}$)       & 1.89 $\pm$ 0.26 \\
$M$\,sin\,$i$ ($M_\earth$) & 4.15\,$\pm$\,0.58\\
$a$ (AU)                & 0.0500\,$\pm$\,0.0007\\
$N_\mathrm{obs}$ (binned) & 86 \\
Median binned uncertainty (\mse) & 0.99\\
Assumed jitter (\mse) & 1.50 \\
$\sigma$ (\mse) & 1.74 \\
$\sqrt{\chi^2_\nu}$  & 0.97 \\
\enddata
\end{deluxetable}

Figure~\ref{fig:ts_pg}b/f shows the RV measurements after subtracting 
the long-period signal.  Trends in the time series are no longer apparent and the 
4.647\,d signal and its alias at 1.270\,d are significantly strengthened in the periodogram.  
Figure~\ref{fig:ts_pg}c/g shows the RVs after subtracting both long and short period signals.  
The value of $\sigma$ is 1.74\,\ms and the periodogram appears nearly featureless, 
with no periods remaining having significant power.  

\begin{figure*}
\epsscale{1.15}
\plotone{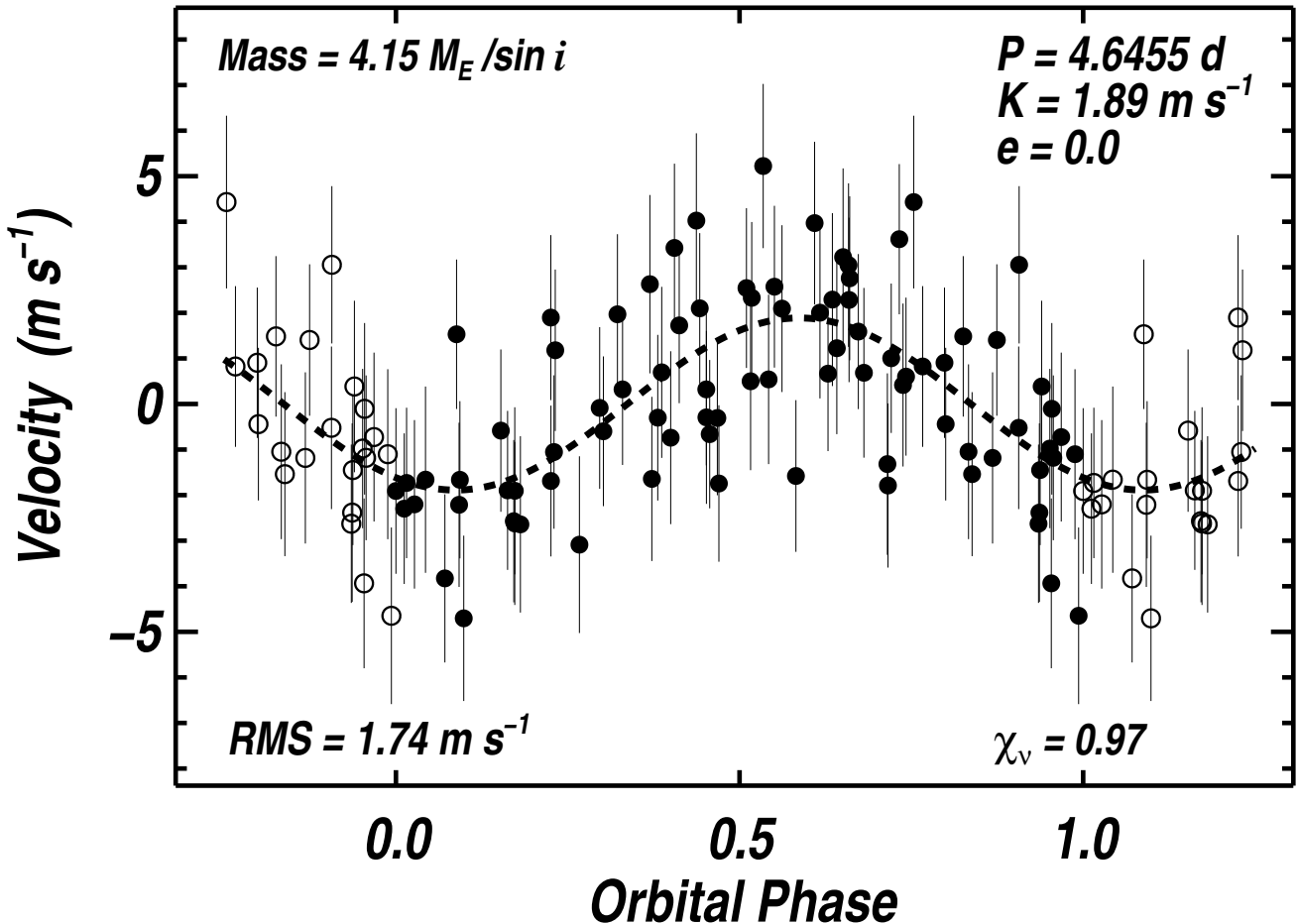}
\caption{Single-planet model for the radial velocities of HD\,156668, 
       as measured by Keck-HIRES.
       The dashed line shows the best-fit circular orbital solution representing a 4.15\,\mearth (minimum mass)
       planet in a 4.6455\,d orbit.
       Filled circles represent phased, binned, and high-pass filtered velocities, 
       while the open circles represent the same velocities wrapped one orbital phase.
       The error bars show the quadrature sum of measurement uncertainties and 1.5\,\ms jitter.}
\label{fig:phased}
\end{figure*}

The phased orbital solution for HD\,156668\,b is shown in Figure~\ref{fig:phased}.  
The Doppler semi-amplitude of $K$\,=\,1.89\,\ms is extremely low and is nearly equal 
to the 1.80\,\ms typical error for single measurements (including jitter). 
The resulting minimum mass of \msinie\,=\,4.15\,\mearth is also extremely small, 
and is the second lowest reported to date using the RV technique.
We adopted a circular orbit because the uncertainty on the eccentricity ($e$\,=\,0.20\,$\pm$\,0.17) 
is significant when that parameter is allowed to float in the two-planet fit.  
Further, the best-fit eccentric orbit model did not show a statistically significant improvement in $\chi_{\nu}$ 
compared to the circular orbit model.

The Keplerian parameter uncertainties were derived using a Monte Carlo method \citep{Marcy05}.  
The uncertainty estimates on \msini and $a$ account for the uncertainty from $M_{\star}$.
With five years of observations of this short-period planet, 
the error on $P$ is quite small (one part in 4000).

\section{The Null Hypothesis}
\label{sec:null}

In this section we explicitly consider the null hypothesis.  
That is, we consider whether the $P$\,=\,4.6455\,d signal we identified as due to a low-mass
planet in Keplerian motion could be due to another source.
Such an analysis is motivated by the extremely low amplitude signal ($K$\,=\,1.89\,\mse) 
and because of the use of high-pass filtering.

\subsection{False Alarm Probabilities}
\label{sec:fap}
We considered the possibility that the $P$\,=\,4.6455\,d signal arose from the chance arrangement of 
random, statistically independent errors in the RVs 
by computing FAPs for several fits to the data \citep{Marcy05,Cumming04,Howard09b,Howard09a}.
These FAPs compare the measured data to 
1000 scrambled data sets drawn randomly with replacement from unscrambled data.  
For each data set we compare a best-fit Keplerian model to 
the null hypothesis (a linear fit to the data) by computing   
$\Delta\chi^2_{\nu}$\,=\,$\chi^2_{\mathrm{lin,}\nu}$\,$-$\,$\chi^2_{\mathrm{Kep,}\nu}$, 
where $\chi^2_{\mathrm{lin,}\nu}$ and $\chi^2_{\mathrm{Kep,}\nu}$ 
are the values of $\chi^2_{\nu}$ for linear and Keplerian fits to the data, respectively.   
The $\Delta\chi^2_{\nu}$ statistic measures the improvement in the fit of a Keplerian model 
compared to a linear model of the same data.  
The FAP is the fraction of scrambled data sets that have a larger value of $\Delta\chi^2_{\nu}$ 
than for the unscrambled data set.  
That is, the FAP measures the fraction of scrambled data sets where the improvement in $\Delta\chi^2_{\nu}$ from a 
best-fit Keplerian model over a linear model is greater than the improvement of 
a Keplerian model over a linear model for the actual measured velocities.
We use $\Delta\chi^2_{\nu}$ as the goodness-of-fit statistic, 
instead of other measures such as $\chi_{\nu}$ for a Keplerian fit, 
to account for the fact that the scrambled data sets, 
drawn from the original velocities \textit{with replacement}, 
have different variances, which sometimes artificially improve the fit quality 
(i.e.\  some scrambled data sets contain fewer outlier velocities and have lower rms).  
Note that this FAP does not measure the probability of 
non-planetary sources of true velocity variation masquerading as a planetary signature.  


We computed an FAP for the ``two-planet model'' (with the long-period planet acting as a high-pass filter) by 
comparing $\Delta\chi^2_{\nu}$ for the actual data with 
$\Delta\chi^2_{\nu}$ for two-planet fits to scrambled sets of the high-pass filtered data 
with the unscrambled long-period signal added back in.
This FAP has a value of 0.3\% (3/1000) and tests whether the statistical significance 
of the short-period signal is somehow enhanced by simultaneously fitting for a long-period signal.

We also computed the FAP for a 1-planet fit to the high-pass filtered data. 
We compared $\Delta\chi^2_{\nu}$ for the unscrambled, filtered data set and scrambled versions thereof.  
This FAP tests whether the short-period signal is a statistical fluctuation, 
while assuming that the long-period signal is real (but still of unknown origin).  
We found an FAP of 0.4\% (4/1000) for this scenario (with eccentricity unrestricted in the fits to scrambled data sets).
We note that \textit{all} four of the false alarm solutions had best-fit parameters that appeared unphysical: 
high eccentricity ($e$\,$>$\,0.5) and short period ($P$\,$<$\,10\,d).  
(We regard them as unphysical because short-period planets are almost universally in near circular orbits.)
Such spurious, high eccentricity solutions often appear as the best-fit solution 
to low rms RV data lacking a coherent signal.  
When we restricted the fits of scrambled data sets to circular orbits (as we did with the actual measurements), 
the FAP dropped to $<$\,0.1\% (0/1000).

We conclude that the $P$\,=\,4.6455\,d signal is statistically significant.  
Considering all of the FAP tests, 
we estimate that the probability that this signal arose just due to random errors is less than 1\% 
and probably on the order of 0.1\%.

\subsection{Photometric Observations}
\label{sec:photometry}

In addition to the Keck RVs, we obtained contemporaneous
Str\"omgren $b$ and $y$ photometric measurements with the T10 0.80\,m automatic 
photometric telescope (APT) at Fairborn Observatory in Arizona.  This APT is 
identical to the T8 APT that was used to acquire photometric observations 
of HD\,7924 in the first paper of this series \citep{Howard09a}.  (That paper 
mistakenly identified the APT used as the T12 0.8\,m APT.)  Our observing 
procedures and data reduction procedures in this paper are identical to those 
described in \citet{Howard09a}. 

Our three comparison stars A, B, and C for HD\,156668 (star D) were HD\,158974 
($V=5.63$, $B-V=0.96$, G8~III), HD\,155092 ($V=7.07$, $B-V=0.42$, F2), and 
HD\,156536 ($V=7.51$, $B-V=0.42$, F3~IV), respectively.  The T10 APT acquired 
477 good observations of this quartet of stars covering the 2007, 2008, 
and 2009 observing seasons.  Comparison star A, HD\,158974, exhibits
low-amplitude brightness variability of 5\,mmag with a period of 152 days.  
The $C-B$ differential magnitudes have a standard deviation of 1.7\,mmag,
indicating that both stars are constant to the level of our measurement 
precision.  Therefore, we created differential magnitudes of HD\,156668 by 
averaging the $D-B$ and $D-C$ differential magnitudes into a single 
$D-(B+C)/2$ differential magnitude to be used for our analyses.  We also 
combined our $b$ and $y$ observations into a single $(b+y)/2$ passband to 
improve the precision further.

\begin{figure}
\epsscale{1.1}
\plotone{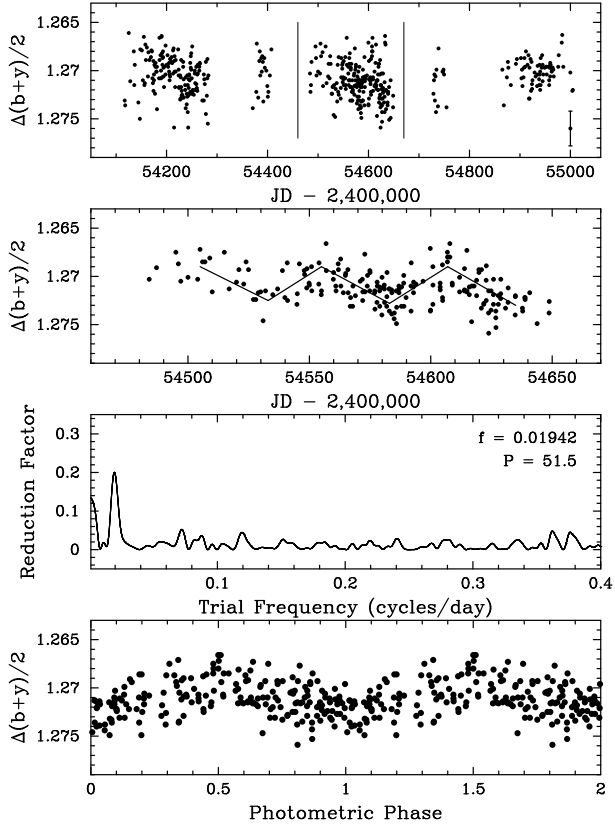}
\caption{$Top$:  The 477 $D-(B+C)/2$ photometric observations of HD\,156668
in the $(b+y)/2$ passband, acquired with the T10 0.8~m APT during the 2007,
2008, and 2009 observing seasons. $Second~panel$:  The portion of the
2008 observing season set off by vertical bars in the top panel shows the
most coherent brightness variability due to cool star spots carried across
the disk of the star by its rotation.  $Third~panel$:  Frequency spectrum
of the observations in the second panel gives a best period of 51.5 days.
$Bottom~Panel$:  Plot of the data from panel two phased with the
51.5-day period reveals coherent variability with a peak-to-peak
brightness amplitude 0.0023 mag.}
\label{fig:phot1}
\end{figure}

These $D-(B+C)/2$ magnitudes in the $(b+y)/2$ passband are shown in the top 
panel of Figure~\ref{fig:phot1}.  The gaps following the longer runs of data are due to
the necessity of shutting down APT operations during the summer rainy
season in Arizona.  The gaps following the shorter data groups are the normal
seasonal gaps for HD\,156668.  A summary of the photometric observations
is given in Table~\ref{tab:photometry}.  Column~5 indicates that the seasonal mean magnitudes
vary over a range of $\sim$1\,mmag, which is typical for solar-type stars 
with modest chromospheric activity \citep[see, e.g.,][]{Lockwood07}.

\begin{deluxetable*}{ccccc}
\tablenum{1}
\tablewidth{0pt}
\tablecaption{Summary of Photometric Observations of HD\,156668\label{tab:photometry}}
\tablehead{
\colhead{Observing} & \colhead{} & \colhead{Date Range} & \colhead{Sigma} & \colhead{Seasonal Mean} \\
\colhead{Season} & \colhead{$N_{obs}$} & \colhead{(HJD $-$ 2,400,000)} & \colhead{(mag)} & \colhead{(mag)} \\
\colhead{(1)} & \colhead{(2)} & \colhead{(3)} & \colhead{(4)} & \colhead{(5)} 
}
\startdata
 2007 & 181 & 54117--54407 & 0.00195 & $1.27046\pm0.00014$ \\
 2008 & 225 & 54484--54754 & 0.00188 & $1.27121\pm0.00013$ \\
 2009 &  71 & 54865--55004 & 0.00131 & $1.27013\pm0.00016$ \\
\enddata
\end{deluxetable*}

The vertical bars in the top panel of Figure~\ref{fig:phot1} encompass the portion of the
$D-(B+C)/2$ light curve that most clearly exhibits coherent variability
that might be due to rotational modulation of spots on the star's 
photosphere \citep[see, e.g.,][]{Henry1995}. That section of the light curve 
is replotted in the second panel.  Straight line segments approximate the 
2.5 cycles of the purported brightness variability.

Plotted in the third panel of Figure~\ref{fig:phot1} is the frequency spectrum of the
observations from panel two, showing a clear periodicity of 51.5\,d. 
In the bottom panel, the same observations are phased to the 51.5\,d 
period and a time of minimum computed with a least-squares sine fit to 
the observations.  The sine fit also gives a peak-to-peak amplitude of 
$2.3\pm0.3$\,mmag.

We take the 51.5\,d period to be the rotation period of HD\,156668.  This
is consistent with $v$\,sin\,$i$\,=\,0.5\,km\,s$^{-1}$  
and is very close to the rotation period of 48\,d predicted from its level
of chromospheric activity. 

\begin{figure}
\epsscale{1.1}
\plotone{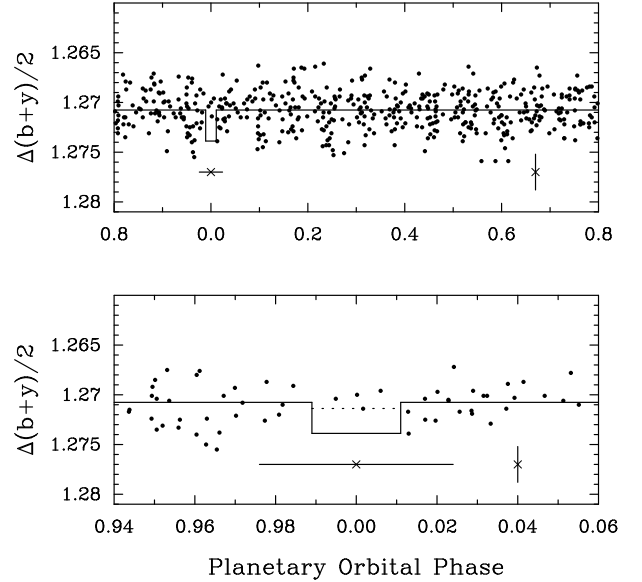}
\caption{$Top~panel$: The 477 APT photometric observations of HD\,156668
plotted modulo the 4.6455\,d period of the RV variations. 
A least-squares sine fit at that period yields a semi-amplitude of only 
0.10\,$\pm$\,0.11\,mmag, providing strong evidence that the velocity
variations arise from a planetary companion.  $Bottom~panel$:  The 
photometric observations of HD\,156668 near the predicted time of transit 
replotted with an expanded scale.  The solid curve shows the predicted 
depth (3.1\,mmag) and duration ($\pm0.011$ phase units) of a central 
transit, computed from the orbital elements, the stellar properties, and a 
planetary composition of pure hydrogen.  The uncertainty in the predicted 
transit time is shown by the error bar under the transit window.  The
dotted line across the transit window shows the expected depth (0.6\,mmag) 
for a planet composed of water.  Transits of a pure hydrogen
planet are essentially ruled out.  Additional observations are required to
rule out other planetary compositions.}
\label{fig:phot2}
\end{figure}

The APT photometry is also useful for confirming that observed
RV variations are due to a planetary companion and not stellar
surface activity. \citet{Queloz2001} and \citet{Paulson2005} have demonstrated 
how rotational modulation in the visibility of star spots on active stars 
can result in periodic RV variations and their misinterpretation.
All 477 photometric observations of HD\,156668 are plotted in the top panel 
of Figure~\ref{fig:phot2}, phased with the 4.6455\,d RV period and a time 
of mid-transit computed from the orbital elements.  A least-squares sine 
fit on that period gives a semi-amplitude of only $0.10\pm0.11$\,mmag.  
This absence of detectable rotational modulation of surface activity to 
high precision on the RV period provides strong evidence that
the RV variations arise from a super-Earth companion.

The bottom panel of Figure~\ref{fig:phot2} replots the photometric observations that lie 
near the predicted time of transit.  The solid curve shows the predicted
depth (3.1\,mmag) and duration ($\pm0.011$ phase units) of a central
transit, computed from the orbital elements and stellar properties with
an assumed planetary composition of pure hydrogen.  That such shallow
transits can be detected with the APTs has been shown by \citet{Sato05},
who used the T11 APT to discover the 3\,mmag transits of HD\,149026b.
The dotted line across the transit window corresponds to the expected
depth (0.6\,mag) of a hypothetical planet composed of entirely water.  Our observations show 
that transits with a depth of 3.1\,mag probably due not occur, essentially 
ruling out transits of a hydrogen planet.  Additional precise photometry is 
required to rule out other planetary compositions.

\subsection{Chromospheric Activity}
\label{sec:chromospheric}

RV planet searches measure the shifts of the centroids of thousands of stellar lines.  
The shapes of individual lines are determined in large part by 
Doppler broadening from the disk-averaged velocity field of the star's surface.
Spots are magnetically-controlled regions of the stellar photosphere characterized 
by temperatures $\sim$1000\,K lower than unaffected regions.  
As they rotate across the stellar surface, spots contribute less flux to 
particular parts of each absorption line profile---less flux on the blue side of lines for spots on 
the approaching limb and less flux on the red side for spots on the receding limb---thereby 
distorting the average line profile, shifting the centroid, and causing an apparent Doppler shift of the star.  

False positive signals of this type tend to occur around chromospherically active stars 
\citep{Queloz2001,Paulson2005}.  
The spurious RV signals are coherent over typical spot lifetimes (weeks to months) and have periods similar 
to the stellar rotation period.
In \S\,\ref{sec:photometry} we showed that the 4.6455\,d signal is not seen in APT photometry, 
ruling out rotationally modulated spots as the source of the RV periodicity.
Here, we use the chromospheric indices $S_\mathrm{HK}$ and \lrphk\ to strengthen that conclusion further.
These indices measure the level of stellar chromospheric activity,
which in turn is strongly correlated with the magnetic activity of the stellar photosphere.  

Some stars also show long-term chromospheric activity cycles as the average number of spots 
rises and falls with the solar cycle, typically with a timescale of $\sim$10\,yr.  
These cycles are sometimes detected as apparent RV shifts and incorrectly interpreted as long-period planets.

\begin{figure}
\epsscale{1.20}
\plotone{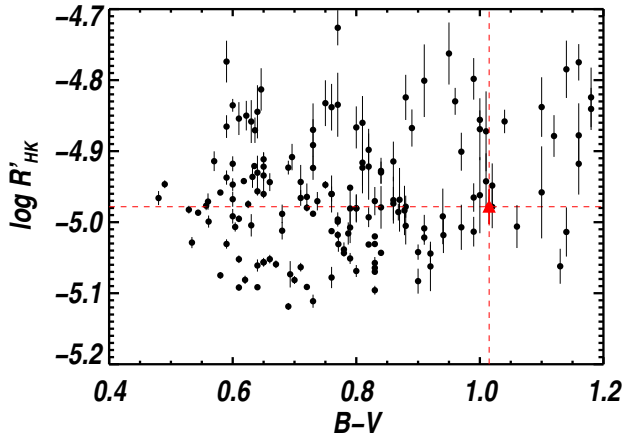}
\caption{Plot of \lrphk\ as a function of $B-V$ color for all G and K stars in the Eta-Earth survey 
    with $B-V$\,$<$\,1.2.
    HD\,156668 (large red triangle highlighted by dashed red lines) 
    is among the least chromospherically active 
    Eta-Earth stars with similar $B-V$.}
\label{fig:rphk}
\end{figure}

Figure~\ref{fig:rphk} shows the average value of \lrphk\ for HD\,156668 and the other 
stars in the Eta-Earth Survey with $B-V$\,$<$\,1.2.  
HD\,156668 is among the least chromospherically active stars of similar spectral type 
in the Eta-Earth Survey.  
This is consistent with our limits on photometric variability from APT measurements (\S\,\ref{sec:photometry}) 
and the small RV residuals to the Keplerian fit.  

The values of $S_\mathrm{HK}$ derived from each Keck-HIRES spectrum are listed in 
Table~\ref{tab:keck_vels} and plotted as a time series in Figure~\ref{fig:ts_pg}d.  
A Lomb-Scargle periodogram of the data is plotted in Figure~\ref{fig:ts_pg}h.  
Importantly, the periodogram shows negligible power for $P$\,$\lesssim$\,20\,d, 
as expected for a middle-aged star with a predicted rotation period of $\sim$48\,d.  
This lack of power at periods near 4.6\,d strengthens the case for HD\,156668\,b.

On average, $S_\mathrm{HK}$ rises with time (Figure~\ref{fig:ts_pg}d), 
possibly due to the observation of a partial solar magnetic cycle.  
This several year long trend is not observed in the RVs (Figure~\ref{fig:ts_pg}a).  
One feature of the RV time series is also apparent in the $S_\mathrm{HK}$, 
the declining trend in late 2007.  
This apparent correlation raises the possibility that the long-period signal that was filtered out of the RVs 
in \S\,\ref{sec:orbital} is due to stellar activity.
Despite this apparent correlation, the Pearson linear correlation coefficient, $r$\,=\,$+$0.11, 
between the unfiltered RVs and the $S_\mathrm{HK}$ values indicates an insignificant correlation.  
When we subtract a second-order polynomial fit to the $S_\mathrm{HK}$ values, 
the correlation with the RVs is still statistically insignificant, $r$\,=\,$+$0.19.
Thus, the suggestion that stellar activity explains the long-period signal is not well supported 
when the entire data set is considered.

Rotational modulation of spots and other surface features can also be detected in the 
spectral line bisectors \citep{Torres05}.  
However, for this extremely low amplitude signal ($K$\,$<$\,2\,\mse), 
detecting line profile variations at the same or higher precision is not possible.  

\begin{figure}
\epsscale{1.20}
\plotone{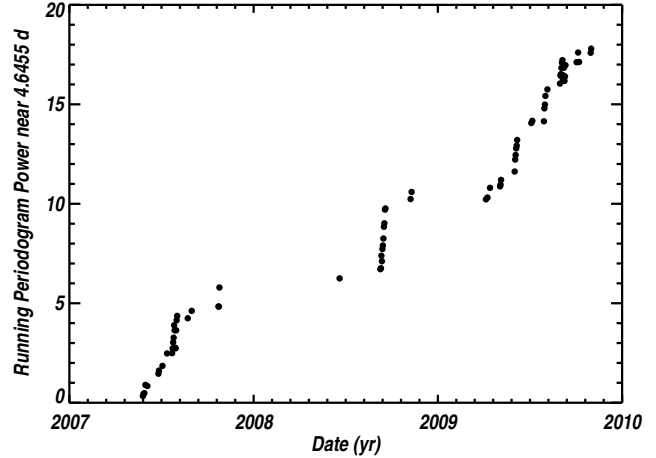}
\caption{Running periodogram power near $P$\,=\,4.6455\,d as a function of time.  
    Each dot represents the maximum periodogram power for the set of 
    all high-pass filtered RVs up to that point in time.  
    Only periods near the orbital period of HD\,156668\,b ($P$\,=\,4.64--4.65\,d) were used to  
    compute each maximum periodogram power value.
    The monotonic rise confirms that the signal was present throughout the observations 
    and points to a dynamical origin.}
\label{fig:running_pergram}
\end{figure}

As a final check, we show in Figure\ \ref{fig:running_pergram} that the RV signal from 
HD\,156668\,b is strictly periodic and present throughout the observations, 
as the clock-like signal from a planet should be.
The plot shows the periodogram power of the planet rising monotonically as additional measurements are taken.  
False periodicities, such as those due to spots, typically exhibit periodicities that are only briefly coherent 
and that may reappear with slightly different periods.  

\subsection{Stroboscopic Alias}
\label{sec:alias}

We interpret the periodogram spike near 1.270\,d (Figure~\ref{fig:ts_pg}) as 
a stroboscopic alias of the orbital period 
(which appears on the periodogram as 4.647\,d due to finite period sampling)
with the sidereal day length (0.997\,d): 1/4.647\,d + 1/1.270\,d = 1/0.997\,d.
We verified this by computing the periodograms of 2000 mock data sets that 
have the same times of observations and errors as the actual RVs.  
Of the 1000 data sets that we injected a 4.647\,d signal, 851 of them showed significant 
periodogram power ($\ge10$) near 1.27\,d.  
Conversely, 0/1000 of the mock data sets with a 1.270\,d signal added showed significant power near
4.647\,d.  

\section{Discussion}
\label{sec:discussion}

We present the detection of HD\,156668\,b, a super-Earth planet with minimum mass 
$M_P\sin i$\,=\,4.15\,$M_\earth$ in a $P$\,=\,4.6455\,d orbit around a K3 dwarf.
We draw on several lines of evidence to support the existence of HD\,156668\,b.  
We showed in \S\,\ref{sec:fap} that the short-period signal is statistically significant.  
This signal is apparent in a fit to the unfiltered RVs, 
and stands out strongly when isolated by high-pass filtering.
The host star is middle-aged and quiet, providing a nearly ideal RV target.
The planet's short-period signature is not seen in photometric observations or in 
chromospheric indices.  
Thus, the evidence strongly points to a planetary interpretation.  

To estimate the orbital parameters of HD\,156668\,b, 
we found it necessary to apply a high-pass filter to the RVs by subtracting the model of a long-period Keplerian.  
This model changes slowly over time and clarifies, but does not artificially enhance, 
the $P$\,=\,4.6455\,d signal of HD\,156668\,b.  
Filtering of this type is common in other areas of time domain astronomy 
(e.g.\ transit photometry, astroseismology) and is uncommon, but not unprecedented, in RV planet detection.
Similar filtering techniques were employed to disentangle the planetary signals of 
Gl\,176\,b \citep{Forveille09} and Gl\,674\,b \citep{Bonfils07}, 
although in those cases the non-planetary signal was clearly due to spots modulated by stellar rotation.
\cite{Queloz09} used a ``pre-whitening'' technique to extract the signatures of two super-Earths
from the complicated RV time series of the significantly more active star, Corot-7.  
Further, many exoplanets are announced with a model that includes a 
linear or second order RV trend, presumably due to a long period orbital companion.  
The source of the long-period signal remains unknown and motivates additional measurements.  
If it is due to a planet, the body has minimum mass $\sim$45\,\mearth and 
orbits with $P_c$\,=\,2.31\,yr and $a_c$\,=\,1.6\,AU, a cold super-Neptune near the ice line.

We see no evidence for transits of HD\,156668\,b down to a level of $\sim$3\,mmag.  
However, given the large \textit{a priori} transit probability of 7\%, 
it is instructive to speculate about the transit signatures of various possible planet compositions.  
Using the models in \cite{Seager2007}, a 4\,\mearth planet composed of pure Fe, MgSiO$_3$, H$_2$O, or H 
would yield planets of radius $R_{\mathrm{pl}}$\,=\,1.2, 1.5, 2.0, and 4.5\,\rearthe, 
producing transits of depth 0.22, 0.35, 0.61, and 3.1\,mmag, respectively.  
These homogeneous planet models are oversimplified, but set the scale for admixtures of those ingredients: 
transits of planets made of solids and water would have depths of $\sim$0.2--0.6\,mmag, 
while transits of a planet with a significant atmosphere could be much deeper.  
For comparison, the transiting super-Earth Corot-7\,b 
has a transit depth of 0.36\,mmag 
implying a radius $R_{\mathrm{pl}}$\,=\,1.68\,\rearth \citep{Leger09}.  
Using the RV-determined mass $M_{\mathrm{pl}}$\,=\,4.8\,\mearth \citep{Queloz09}, 
the bulk density is terrestrial, $\rho_{\mathrm{pl}}$\,=\,5.6\,g\,cm$^{-3}$.  
In contrast, GJ\,1214\,b has $M_{\mathrm{pl}}$\,=\,6.6\,\mearth and $R_{\mathrm{pl}}$\,=\,2.7\rearthe, 
implying $\rho_{\mathrm{pl}}$\,=\,1.9\,g\,cm$^{-3}$, 
intermediate between Earth and the ice giants of the Solar System.
Transits of GJ\,1214\,b are unusually deep (15\,mmag) for a planet of this size 
because it orbits an M dwarf with $R_{\star}$\,=\,0.21\,$R_{\sun}$ \citep{Charbonneau09}.

Several authors (e.g.\ \citealt{Ida04a}, \citealt{Mordasini09}) 
have argued that super-Earths will have insignificant hydrogen atmospheres (by mass) because 
during formation their masses failed to reach a critical mass (typically $\sim$10\,\mearthe) when 
the growth from solid planetesimals is augmented by runaway accretion of gas from the protoplanetary disk.
Smaller atmospheres (up to several per cent by mass) could be produced by degassing during impact accretion 
and geological activity \citep{Elkins-Tanton08,Kite09}.
However, whatever atmosphere is acquired from these processes may be lost to atmospheric escape \citep{Baraffe06,Valencia09}.
Nevertheless, the brief history of exoplanets is replete with observational surprises (hot Jupiters, eccentric orbits, etc.) 
so we consider the observational consequences of an atmosphere.  
\cite{Adams08} find that adding a H/He gas envelope equivalent to 0.2--20\% of the mass of a solid 5\,\mearth 
exoplanet increases the radius 8--110\% above the gas-free value.  
Atmospheres dominated by heavier molecules such as H$_2$O and N$_2$ (as on Earth) 
would swell the planet less for the same atmospheric mass 
because of the higher mean molecular weight and reduced scale height.
Thus, we conclude that the APT photometric observations rule out transits for HD\,156668\,b 
if the radius is dominated by a H/He atmosphere (tens of per cent by mass), 
but do not preclude transits if the atmosphere is less massive or composed of heavier elements.

HD\,156668\,b adds statistical weight to the emerging trends of the properties of 
super-Earths and their hosts.  
Like most other such planets, it orbits a K or M dwarf.  
In contrast to other super-Earth hosts, HD\,156668 has a slightly super-solar metallicity \citep{Howard09a}. 
The rate of multiplicity in systems with super-Earths and Neptune-mass planets 
appears to be much higher than for higher mass planet hosts, 
with HD\,40307 \citep{Mayor09}, GJ\,581 \citep{Mayor09b}, and HD\,69830 \citep{Lovis06}
being the standard examples of multiplicity in low-mass systems.  
The long-period signal seen for HD\,156668 is suggestive of a second low-mass planet in the system. 



HD\,156668\,b pushes the frontier of RV planet discovery to lower masses and smaller Doppler amplitudes.
It is the second lowest minimum mass exoplanet discovered to date by the RV technique, after GJ\,581\,e \citep{Mayor09b}.
With a Doppler semi-amplitude of 1.89\,\mse,
HD\,156668\,b is also only the second exoplanet discovered to date with $K < 2.0$\,\mse.   
GJ\,581\,e is the other with $K$\,=\,1.85\,\ms and \msinie\,=\,1.9\,\mearth \citep{Mayor09b}.
This progress is remarkable: 51\,Peg\,b was discovered \citep{Mayor95} 
with \msinie\,=\,0.47\,\mjup and $K$\,=\,57\,\ms 
while the signal from HD\,156668\,b is smaller by factors of 35 and 30, respectively.  
However, we expect that true Earth analogs---1\,\mearth planets in 1\,AU orbits around G stars---will remain 
permanently out of reach of the Doppler technique and will instead be discovered by transit photometry, astrometry, 
and microlensing.  
The Earth imparts a $K$\,=\,0.09\,\ms signal on the Sun, a factor of 20 smaller than HD\,156668\,b.  
Our Keck/HIRES observations of HD\,156668 take $\sim$5\,min to achieve 1\,\ms precision; 
scaling to 0.1\,\ms precision requires 100 times the photons and impractically long integration times.
(A handful of exceptionally bright nearby stars such as $\alpha$ Cen A/B may be an exception \citep{Guedes08}.)
Moreover, stellar jitter, even when averaged over with long integrations, 
likely limits Doppler precision to a few tenths of a \ms for the most quiet stars.  
Longer period planets suffer from the additional disadvantage that only a few orbits transpire during an observational 
campaign (compared with $\sim$100 orbits for the discoveries of HD\,155668\,b and GJ\,581\,e), 
reducing their detectability by clock-like coherence.  

Nevertheless, the future of the Doppler technique is bright 
and we expect that it will continue to play a prominent role in planet searches.
Discoveries of super-Earths and Neptunes from the Eta-Earth Survey will shape our understanding of 
planet formation and migration.  
Doppler work in other domains---long period giant planets, subgiants, multi-planet systems, 
dynamically interacting planets, Rossiter-McLaughlin measurements, etc.---will also continue 
to inform and enrich our knowledge of the rich astrophysics of exoplanets.

\acknowledgments{We thank the many observers who contributed to the velocities reported here.  
We gratefully acknowledge the efforts and dedication of the Keck Observatory staff, 
especially Scott Dahm, Hien Tran, and Grant Hill for support of HIRES 
and Greg Wirth for support of remote observing.  
We thank Heather Knutson, Doug Lin, Shigeru Ida, Jeff Scargle, and Ian Howard for helpful discussions. 
We are grateful to the time assignment committees of the University of California, NASA, and NOAO  
for their generous allocations of observing time.  
Without their long-term commitment to RV monitoring, 
these long-period planets would likely remain unknown.  
We acknowledge R.\ Paul Butler and S.\ S.\ Vogt for many years
of contributing to the data presented here.
A.\,W.\,H.\ gratefully acknowledges support from a Townes Post-doctoral Fellowship 
at the U.\,C.\ Berkeley Space Sciences Laboratory.
G.\,W.\,M.\ acknowledges NASA grant NNX06AH52G.  
G.\,W.\,H.\ acknowledges support from NASA, NSF, Tennessee State University, and
the State of Tennessee through its Centers of Excellence program.
This work made use of the SIMBAD database (operated at CDS, Strasbourg, France), 
NASA's Astrophysics Data System Bibliographic Services, 
and the NASA Star and Exoplanet Database (NStED).
Finally, the authors wish to extend special thanks to those of Hawai`ian ancestry 
on whose sacred mountain of Mauna Kea we are privileged to be guests.  
Without their generous hospitality, the Keck observations presented herein
would not have been possible.}

\bibliographystyle{apj}
\bibliography{hd156668}


\LongTables  
\begin{deluxetable}{cccc}
\tabletypesize{\footnotesize}
\tablecaption{Radial Velocities and $S_{\mathrm{HK}}$ values for HD\,156668
\label{tab:keck_vels}}
\tablewidth{0pt}
\tablehead{
\colhead{}         & \colhead{Radial Velocity}     & \colhead{Uncertainty}  & \colhead{}  \\
\colhead{JD -- 2440000}   & \colhead{(\mse)}  & \colhead{(\mse)}  & \colhead{$S_\mathrm{HK}$}
}
\startdata
 13478.97768 &   -1.04 &    1.03  &          0.202                     \\ 
 13547.90964 &   -4.86 &    0.99  &          0.214                     \\ 
 13604.83890 &   -5.21 &    0.85  &          0.213                     \\ 
 13807.14411 &    3.29 &    1.14  &          0.222                     \\ 
 13932.91866 &    0.87 &    1.00  &          0.229                     \\ 
 13960.91401 &    3.29 &    0.87  &          0.220                     \\ 
 13961.80956 &    0.06 &    0.88  &          0.219                     \\ 
 13981.77064 &    3.87 &    0.94  &          0.231                     \\ 
 13982.87659 &    0.94 &    0.84  &          0.233                     \\ 
 13983.81911 &    4.28 &    0.80  &          0.230                     \\ 
 13984.90602 &    5.74 &    0.85  &          0.231                     \\ 
 14247.01977 &    1.27 &    0.71  &          0.239                     \\ 
 14248.00711 &    1.47 &    1.10  &          0.238                     \\ 
 14249.92154 &    0.79 &    1.14  &          0.235                     \\ 
 14252.01612 &    4.28 &    1.08  &          0.234                     \\ 
 14255.84743 &   -0.27 &    0.76  &          0.238                     \\ 
 14277.79111 &   -3.66 &    1.11  &          0.230                     \\ 
 14278.80991 &   -2.37 &    0.98  &          0.230                     \\ 
 14285.81562 &    0.75 &    1.10  &          0.230                     \\ 
 14294.89361 &    2.08 &    1.16  &          0.226                     \\ 
 14304.95027 &   -0.99 &    0.75  &          0.232                     \\ 
 14305.95179 &   -2.31 &    0.69  &          0.234                     \\ 
 14306.92739 &   -2.31 &    0.70  &          0.232                     \\ 
 14307.97764 &   -0.33 &    0.73  &          0.231                     \\ 
 14308.94701 &    2.07 &    1.00  &          0.229                     \\ 
 14309.94163 &    0.68 &    0.71  &          0.232                     \\ 
 14310.93496 &    0.77 &    0.66  &          0.230                     \\ 
 14311.92755 &   -1.40 &    0.67  &          0.230                     \\ 
 14312.93048 &    1.49 &    0.72  &          0.228                     \\ 
 14313.92787 &    2.74 &    0.68  &          0.228                     \\ 
 14314.96879 &   -2.10 &    1.01  &          0.227                     \\ 
 14335.87372 &   -2.54 &    0.63  &          0.229                     \\ 
 14343.79865 &   -4.20 &    0.89  &          0.219                     \\ 
 14396.70306 &   -3.13 &    0.95  &          0.221                     \\ 
 14397.70527 &   -4.94 &    0.93  &          0.222                     \\ 
 14398.73491 &   -6.92 &    1.11  &          0.219                     \\ 
 14636.95177 &   -3.36 &    1.23  &          0.238                     \\ 
 14717.74461 &    3.34 &    0.99  &          0.254                     \\ 
 14718.89500 &   -0.22 &    0.97  &          0.252                     \\ 
 14719.78911 &   -4.40 &    1.02  &          0.256                     \\ 
 14720.82821 &    2.28 &    0.92  &          0.255                     \\ 
 14721.81321 &    5.53 &    1.00  &          0.255                     \\ 
 14722.75844 &    0.73 &    0.97  &          0.255                     \\ 
 14723.75113 &   -0.66 &    0.90  &          0.255                     \\ 
 14724.77287 &   -2.25 &    0.96  &          0.254                     \\ 
 14725.74632 &    0.03 &    1.03  &          0.252                     \\ 
 14726.80657 &    4.30 &    0.97  &          0.250                     \\ 
 14727.81096 &    1.82 &    0.92  &          0.250                     \\ 
 14777.68651 &    2.56 &    1.05  &          0.236                     \\ 
 14779.69190 &   -4.18 &    1.23  &          0.235                     \\ 
 14927.05569 &   -1.01 &    1.31  &          0.251                     \\ 
 14930.09564 &    2.92 &    1.26  &          0.244                     \\ 
 14935.05529 &    4.30 &    1.19  &          0.237                     \\ 
 14955.10678 &    4.61 &    1.16  &          0.239                     \\ 
 14955.97159 &    0.56 &    1.14  &          0.241                     \\ 
 14957.05565 &   -1.73 &    1.05  &          0.241                     \\ 
 14983.97707 &   -0.72 &    1.09  &          0.241                     \\ 
 14984.96665 &   -2.64 &    1.22  &          0.242                     \\ 
 14985.92186 &    0.69 &    1.13  &          0.239                     \\ 
 14986.99173 &    1.99 &    1.14  &          0.240                     \\ 
 14987.99574 &   -1.07 &    1.20  &          0.242                     \\ 
 14988.89618 &   -2.23 &    1.08  &          0.244                     \\ 
 15016.97485 &   -4.08 &    1.08  &          0.253                     \\ 
 15019.04148 &    0.23 &    1.25  &          0.252                     \\ 
 15041.97022 &   -0.80 &    1.16  &          0.236                     \\ 
 15042.89409 &    2.71 &    1.24  &          0.235                     \\ 
 15043.90466 &   -1.71 &    1.13  &          0.233                     \\ 
 15044.94394 &   -2.75 &    1.00  &          0.233                     \\ 
 15048.86672 &   -2.97 &    1.26  &          0.239                     \\ 
 15073.76736 &   -1.01 &    0.95  &          0.255                     \\ 
 15074.75962 &    1.61 &    0.91  &          0.253                     \\ 
 15075.73804 &    0.05 &    0.66  &          0.252                     \\ 
 15076.74458 &   -2.42 &    0.68  &          0.247                     \\ 
 15077.74489 &   -1.57 &    0.96  &          0.242                     \\ 
 15078.76561 &   -2.64 &    1.00  &          0.239                     \\ 
 15079.73946 &   -2.60 &    0.72  &          0.237                     \\ 
 15080.74369 &   -0.13 &    0.69  &          0.235                     \\ 
 15081.73701 &   -3.35 &    0.69  &          0.234                     \\ 
 15082.72821 &    0.83 &    1.01  &          0.229                     \\ 
 15083.73498 &    1.02 &    0.69  &          0.230                     \\ 
 15084.74428 &    1.19 &    1.00  &          0.228                     \\ 
 15106.76608 &   -2.26 &    1.16  &          0.256                     \\ 
 15109.75425 &   -3.25 &    1.39  &          0.261                     \\ 
 15111.72700 &   -1.94 &    0.79  &          0.263                     \\ 
 15134.69674 &   -0.50 &    0.82  &          0.232                     \\ 
 15135.70259 &   -1.59 &    0.80  &          0.229                     
\enddata
\end{deluxetable}

\enddocument